\documentclass[12pt]{article}  
\newcommand{\comma}{\;\; , \; \; }
\newcommand{\period}{\;\; .}
\newcommand{\eq}{\; = \;}
\newcommand{\sep}{\;\; , \;\;}

\newcommand{\be}{\begin{equation}}
\newcommand{\bd}{\begin{displaymath}}
\newcommand{\ee}{\end{equation}}
\newcommand{\ed}{\end{displaymath}}
\newcommand{\ba}{\begin{eqnarray}}
\newcommand{\ea}{\end{eqnarray}}

\newcommand{\mod}{{\rm mod}\,}

\renewcommand{\i}{{\rm i}}

\newcommand{\e}{{\rm e}}

\renewcommand{\theequation}{\arabic{section}.\arabic{equation}}
\newcommand{\W}{\overline{\cal W}}

\newcommand{\Ws}{{\cal{W}}}

\newcommand{\q}{p}

\newcommand{\ha}{{{\cal H}}_0}
\newcommand{\hb}{{{\cal H}_1}}
\newcommand{\tZ}{{\tilde Z}}
\newcommand{\tW}{{\widetilde W}}
\newcommand{\cV}{{\cal V}}
\newcommand{\sd}{^{\vphantom{\dagger}}  }
\newcommand{\cS}{{\cal S}}

\title{Some remarks on a generalization of the superintegrable 
chiral Potts model}

\author{ R.J. Baxter\\
{\protect \small  Mathematical
Sciences Institute,  The Australian National}\\
{\protect  \small  University,
 Canberra, A.C.T. 0200, Australia, \small e-mail: none }}

\date{\protect \small 16 June 2009}

\begin{document}


\maketitle


 \abstract{The spontaneous magnetization of a two-dimensional lattice 
model can be expressed in terms of the partition function $W$ of 
a system 
with fixed boundary spins and an extra weight dependent on the value 
of a particular central spin. For the superintegrable case of the 
chiral Potts model with cylindrical boundary conditions, $W$
can be expressed in terms of reduced hamiltonians $H$ and a central 
spin operator $S$. We  conjectured in a previous paper that 
$W$ can  be written as a determinant, similar to 
that of the Ising model. Here we generalize this conjecture to 
any Hamiltonians that satisfy a more general Onsager algebra, 
and give a conjecture for the elements of $S$.}


 \vspace{5mm}




 \section{Introduction}
Onsager calculated the partition function 
of the two-dimensional Ising model 
by noting that the two hamiltonians associated with the transfer 
matrices
generated a finite-dimensional  algebra, now known as the 
Onsager algebra. \cite[eqns. 60, 61]{Onsager1944} 
Later, Kaufman showed the problem could be solved by using 
free-fermion 
(i.e. Clifford algebra) operators.\cite{Kaufman1949} This 
method leads naturally to
determinantal expressions, and indeed Kac and Ward showed that 
the partition function
could be expressed combinatorially as a 
determinant.\cite{KacWard1952},  while
Hurst and Green\cite{HurstGreen}  wrote it as  a pfaffian  (the 
square root of an anti-symmetric 
determinant).  Later it was realized that the Ising model could be 
expressed as a 
dimer problem, giving a direct combinatorial solution in terms of 
pfaffians.\cite{Kasteleyn61,TemperleyFisher,Fisher,Kasteleyn63}

The calculation of the spontaneous magnetization ${\cal M}_0$
is a more difficult problem. Onsager
announced his and Kaufman's result for the ${\cal M}_0$  in 
1949.\cite{Onsager1949} The first 
published proof was by Yang in 1952.\cite{Yang1952} Then in 1963, 
Montroll, Potts and Ward \cite{MPW1963} showed that this problem 
could also be solved
combinatorially in terms of determinants.
To this, one begins by writing ${\cal M}_0$ as 
\be {\cal  M}_0 \eq W/Z \comma \ee
where $W, Z$ are  two partition 
functions (with open, fixed spin boundary conditions). $Z$
is the usual partition function, while $W$ is the partition 
function with an extra weight $\sigma_0$. Here $\sigma_0$ is the spin
on a site $0$ deep inside the lattice. In \cite{MPW1963} $Z, W$ 
are evaluated as determinants.


Like the Ising model, the general solvable  $N$-state chiral 
Potts model is
a  solvable model. It has $N-1$ single-site order parameters
 (spontaneous 
magnetizations)  ${\cal M}_r$, where $r= 1, \ldots, N-1$.
Its transfer matrices satisfy the star-triangle 
relation.\cite{BaxPerkAu-Yang}
It is, however, much more difficult mathematically. Its 
free energy
(the logarithm of the partition function) was calculated 
in 1988,\cite{RJB1988}
but it was not until 2005 that ${\cal M}_r$   was 
calculated
by solving certain functional relations derived from the 
star-triangle 
relation.\cite{Baxter2005b} 
The calculation verified a long-standing conjecture of 
Albertini {\it et al}.\cite {Albertini1989} 

The superintegrable chiral Potts model is a special case 
of the general solvable  
chiral Potts model. It has the same  order parameters, so 
to obtain ${\cal M}_r$ 
for the general model it would be sufficient to obtain it 
for the superintegrable 
case.

Further,  the  superintegrable case has mathematical
properties quite similar to those of the Ising model.  The
hamiltonians ${\cal H}_0, {\cal H}_1$ associated with
the transfer matrices also satisfy
the Onsager algebra. If one imposes cylindrical boundary 
conditions, with fixed-spin open boundary conditions on the 
top and  bottom of the lattice, then we show in section 2 
that $Z   =  u^{\dagger} D U u$, where the vectors 
$u^{\dagger}, u$ are determined by the bottom and top
boundary conditions, and $D, U$ can be taken to be 
exponentials of the hamiltonian 
${\cal H} = {\cal H}_0 +k' {\cal H}_1$ that commutes with 
the transfer matrix.
Also, $W = W(r) =   u^{\dagger} D {\cal S}_r U u$, where
the matrix ${\cal S}_r$ arises from the extra weight
factor $\omega^{r \zeta}$ in eqn. (\ref{defWar}). There 
is a reduced representation in which $D, U$ are direct 
products of two-by-two matrices, as in the Ising model, and 
one can define a reduced form $S_{PQ}$ of the matrix  
${\cal S}_r$ by (\ref{redS}).

We recently conjectured\cite{Baxter2008} that  $W(r)$ can be 
written  as a determinant. As yet we have neither proved 
this conjecture,  nor used it to obtain ${\cal M}_r$, but 
numerical studies strongly  suggest that both the  
conjecture, and the resulting formula for ${\cal M}_r$, are 
correct.

Here we obtain commutation relations 
for $S_{PQ}$ in terms of the reduced 
hamiltonians $H_0, H_1$. We generalize the problem to
one in which $H_0, H_1$ satisfy a quite general 
Onsager algebra, not just that of the  superintegrable chiral Potts 
model.

The commutation relations appear to determine
$S_{PQ}$. We  conjecture their solution and the resulting
determinantal form of $W(r)$. Our expectation  is that these 
generalized conjectures will be easier to establish
than the previous particular one.
 
 \section{Partition function}
 \setcounter{equation}{0}


\setcounter{equation}{0}
\subsection*{Definition}

We use the notation of Ref.\cite{Baxter2008} 
 and define the $N$-state chiral Potts on the square lattice $\cal L$, 
rotated through 
$45^{\circ}$, with $M+1$ horizontal rows, each containing  $L$ spins, 
as in Fig. 1. 


 \setlength{\unitlength}{1pt}
 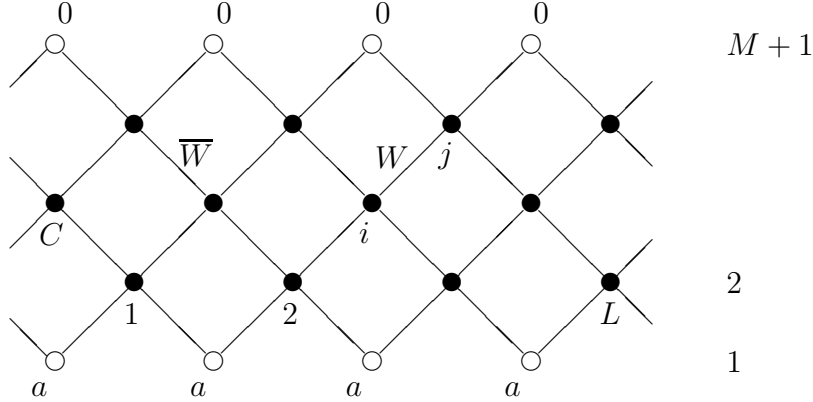
\begin{figure}[hbt]
 \begin{picture}(400,160) (-23,17)

 \put (43,43) {\line(1,1) {74}}
 
 \put (123,63) {\line(1,1) {54}}
 \put (183,63) {\line(1,1) {54}}
 \put (243,63) {\line(1,1) {43}}
 \put (63,3) {\line(1,1) {54}}
 \put (123,3) {\line(1,1) {54}}
 \put (183,3) {\line(1,1) {54}}
 \put (243,3) {\line(1,1) {43}}
 \put (43,104) {\line(1,1) {14}}
 
 \put (63,117) {\line(1,-1) {54}}
 \put (123,117) {\line(1,-1) {54}}
 \put (183,117) {\line(1,-1) {54}}
 \put (243,117) {\line(1,-1) {43}}
   
 \put (43,77) {\line(1,-1) {74}}
 \put (123,57) {\line(1,-1) {54}}
 \put (183,57) {\line(1,-1) {54}}
 \put (243,57) {\line(1,-1) {43}}
  \put (43,16) {\line(1,-1) {14}}

 \multiput(60,0)(60,0){4}{\circle{7}}
 \multiput(60,60)(60,0){4}{\circle*{7}}
 \multiput(60,120)(60,0){4}{\circle{7}}
 \multiput(90,30)(60,0){4}{\circle*{7}}
 \multiput(90,90)(60,0){4}{\circle*{7}}

   \put (51,-13) {$a$}
 \put (111,-13) {$a$}
 \put (170,-13) {$a$}
 \put (230,-13) {$a$}
 
   \put (86,14) {$1$}
   \put (146,14) {$2$}
   \put (266,14) {$L$}

   \put (61,128) {$0$}
    \put (121,128) {$0$}
   \put (181,128) {$0$}
   \put (241,128) {$0$}

 \put (54,44) {$C$}

 \put (175,45) {$i$}
 \put (205,75) {$j$}
  \put (181,73) {$W$}
  
  \put (314,-4) {$1$}
  \put (314,26) {$2$}
  \put (314,116) {$M+1$}
 \put (107,74) {$\overline{W}$}
 
 \end{picture}
 \vspace{1.5cm}
 \caption{\footnotesize The square lattice $\cal L$  turned 
 through $45^{\circ}$.}
 
  \label{sqlatt45}
 \end{figure}


We impose cylindrical boundary 
 conditions, so that the last column $L$ is followed by the first 
column 
 1. At each site  $i$ there is a spin $\sigma_i$, taking the values
 $0, 1, \dots, N-1$. The spins in the bottom row are fixed to have 
 value $a$, those in the top row to have value 0. Adjacent spins  
 $\sigma_i, \sigma_j$ on southwest to northeast edges (with $i$ below 
 $j$) interact with Boltzmann weight ${\W}(\sigma_i - \sigma_j)$;  
 those on southeast to northwest edges with weight 
 $\W(\sigma_i - \sigma_j)$.

   The partition function, which depends on $a$,  is 
   \be Z_a  \eq \sum_{\sigma} \prod_{\langle i,j \rangle} 
  \Ws(\sigma_i - \sigma_j )  \prod_{\langle i,j \rangle} 
  \W(\sigma_i - \sigma_j ) \comma \ee
   the products being over all edges of the two types. The sum is 
  over all values of all the free spins.


  To define the order parameter, we select some inner site $C$ of 
  $\cal L$, say the first site of row $j+1$. Then there are $j$ 
rows of edges below
  $C$ and $M-j$ above. Let $\zeta$ be the spin on site $C$ and 
define
  \be \label{defWar}
W_a (r) \eq \sum_{\sigma} \omega^{r \zeta} 
\prod_{\langle i,j \rangle} 
  \Ws(\sigma_i - \sigma_j )  \prod_{\langle i,j \rangle} 
  \W(\sigma_i - \sigma_j ) \comma \ee
where
 \be \omega \eq \e^{2 \pi \i/N} 
\sep 0  \leq r \leq N\period \ee
   Then the  order parameter is
  \be  \label{defM}
  {\cal M}_r \eq W_0 (r)/Z_0  \comma \ee
 evaluated  in the limit when $L, j, M-j \rightarrow \infty$.

\subsection*{Transfer matrices and hamiltonians}
As in Ref. \cite{Baxter2008}, we define a vector $u_a$, of 
dimension $N^L$, 
with entries
  \ba \label{defua}
   ( u_a)_{\sigma}  & = & 1\;\;  {\rm if }  \;\; \sigma_1 = 
   \cdots  = \sigma_L = a \comma \nonumber \\
     & = & 0 \; \; {\rm otherwise} \period \ea
     We also define a diagonal  $N^L$ by  $N^L$ matrix ${\cS}_r$
with  elements
\be \label{defS}
({\cS}_r)_{\sigma,\sigma'} \eq \omega^{ r \, \sigma_1} 
\prod_{j=1}^L  \delta (\sigma_j, \sigma'_j) \period \ee
We take $0 \leq r \leq N$.

        Let $T$ be the $N^L$ by $N^L$ transfer matrix, defined 
as in
    \cite{Baxter2008},  let $j$ be the number of rows below $C$,
    $M-j$ the number above, and set
    \be \label{DU}
    D = T^j \sep U= T^{M-j} \period \ee
    Then in the usual way, it follows that
    \be \label{defZW}
    Z_a \eq   u_a^{\dagger} \, D U  u_0 \sep     W_a(r)  \eq   
u_a^{\dagger} \, D {\cS}_r  \, U   
    u_0  \comma \ee
    
    The transfer matrix $T$ commutes with a hamiltonian $\cal H$. 
For simplicity,
    we replace the definitions (\ref{DU}) by
   \be \label{DUH}
    D = \e^{-\alpha {\cal H}}  \sep U=   \e^{-\beta {\cal H}}  
\period \ee
    For the ferromagnetic model,  we expect  ${\cal M}_r$ to be 
 unchanged
     if we now define it
    by  (\ref{DUH}), (\ref{defZW}),  (\ref{defM}) and take the 
limit 
    $\alpha, \beta, L \rightarrow + \infty$.     





\subsection*{Superintegrable case}


 Let 
\be \omega = \exp^{2 \pi \i/N}  \ee
and, as in \cite{Albertini1989}, define $N^L$ by $N^L$ 
matrices $Z_j, X_j$  by
\bd
\left( Z_j \right)_{\sigma, \sigma'} = \omega^{\sigma_j} \, 
\prod_{m=1}^L \delta(\sigma_m,\sigma'_m) \comma \ed
\be  \label{ZX}
\left( X_j \right)_{\sigma, \sigma'} =  
\delta(\sigma_j,\sigma'_j+1)
{\prod_{n=1}^L}^{\! \raisebox{-10pt}{*}} 
\delta(\sigma_n,\sigma'_n) 
 \comma \ee
the $*$ on the last product indicating that that it excludes
the case $n=j$. Then from (\ref{defS})
\be 
\label{defS2}
 {\cS}_r = Z_1^{\, r} \period \ee

For the general solvable chiral Potts model, the hamiltonian 
$\cal H$ is given by Albertini {\it et al} \cite{Albertini1989}  
as a 
linear combination of the matrices $Z_j^nZ_{j+1}^{-n}$ and
of $X_j^N$. For the superintegrable case (in their notation
$\phi = \overline{\phi}  = \pi/2$)  this becomes (writing their 
$\lambda$ as $k'$)
\be    \label{defH}
{\cal H} =  {\cal H}_0 +  k' {\cal H}_1    \comma \ee
where
\ba \label{defhh}
{\cal H}_0  & = &  - 2 \sum_{j=1}^L \sum_{n=1}^{N-1} 
\frac{Z_j^nZ_{j+1}^{-n}}{1-\omega^{-n} } \comma \nonumber \\
{\cal H}_1  & = & - 2 \sum_{j=1}^L \sum_{n=1}^{N-1} 
\frac{X_j^n}{1-\omega^{-n} } \period \ea
The    $k'$ in (\ref{defH}) is a ``temperature-like'' parameter, 
satisfying 
\be 0 <  k' < 1 \ee
in the ferromagnetic regime, being small at low temperatures,
and tending towards one  as the system becomes critical.





\subsubsection*{Onsager algebra}


These hamiltonians generate the ``Onsager 
algebra'' [1, eqns. 60,61] and \cite{AuYang1988,AuYangPerk1989,
Davies1990}.
    Define
 \be \label{defA}
A_0 = - 2 \, {\cal H}_1 /N \sep A_1 = 2 \, {\cal H}_0/N 
\comma \ee
  Then there are  two sets of matrices $A_m, G_n$ such that
  \bd
  [A_m, A_n]  \eq 4 G_{m-n}  \comma \ed
  \be  \label{Onsalg}
[G_m,A_n]  \eq    2  A_{m+n}  - 2 A_{n-m} 
\sep [G_m, G_n] \eq 0 \comma \ee
  for all integers $m, n$. 
  
The matrices ${\cal H}_0$,  ${\cal H}_1$ have a highly 
degenerate eigenvalue 
structure. Note that
\be \label{formula}
-  \sum_{n=1}^{N-1} \frac{2 \omega^{k n} }{1-\omega^{-n}} \eq
2 k + 1-N   \sep 0 \leq k < N   \ee
so the LHS is a ``sawtooth'' function, periodic of period $N$, 
linear from $k= 0$
 to $k=N-1$. 
 
 The matrices $Z_j$ are diagonal, and $Z_j^n Z_{j+1}^{-n}$ 
has entries
 $\omega^{n(\sigma_j - \sigma_{j+1})}$. It follows that the 
diagonal elements
 of  ${\cal H}_0$ are of the form
 \be \label{allowed}
 L(1-N) + 2 m N \comma \ee
 where $m$ is an integer and 
 \bd 
 0  \leq m \leq L (N-1)/N \period \ed
There is a similarity transformation that takes $X_j$ to 
$Z_j$. It follows that
the eigenvalues of  ${\cal H}_1$ are also integers of the 
form (\ref{allowed}),
though with different degeneracies from those of ${\cal H}_0$.

\subsubsection*{Commutators with ${\cS}_r$}

To evaluate the matrix elements (2.8), we look at the matrices 
formed by setting
$C_1 = {\cS}_r$ (for a given value of $r$) and then looking at 
the sequence of $C_m$
generated by successively forming  the commutators
$[\ha,C_m]$ and   $[\hb,C_m]$.


It is convenient to define linear operators 
$f_0, f_1$ by
\be \label{deff01}
f_0(C) = \frac{[\ha,C]}{2 N} \sep  f_1(C) =
\frac{ [\hb,C] + 2 r C}{2 N} \ee
for any $N^L$-dimensional 
matrix $C$.




We first note from (\ref{defS2}), (\ref{defH}) that $S_r$, 
${\cal H}_0$ are diagonal
matrices, so $S_r$ commutes with ${\cal H}_0$, so
\be \label{reln1}
f_0(C_1) =  0 \period \ee

We 
can therefore start by forming all 
the linearly independent commutators with ${\cal  H}_1$. If 
we define
\bd C_2 = f_1(C_1) \comma \ed
then we prove in Appendix A  that
\be \label{eqforC2}
f_1(C_2)   = C_2  \ee
so we now have two matrices $C_1, C_2$. They are in general  
linearly  independent.
 
 We have proceeded by performing numerical experiments for 
small $N, L$ and now report our observations. 
 
The next step is to  form all possible commutators with  
 ${\cal  H}_0$. This
leads us to define two more matrices:
 \bd C_3 = f_0(C_2)  \sep C_4 = f_0(C_3)   \comma \ed
 and we find that
 \be \label{reln3}    f_0(C_4)=  C_3 \period \ee
So at this stage we have four matrices, satisfying the three relations
(\ref{reln1}),(\ref{eqforC2}), (\ref{reln3}).

Now we commute with ${\cal  H}_1$, defining four new matrices:
 \bd C_5 = f_1(C_3) \sep C_6 = f_1(C_4) \comma \ed
  \bd C_7 = f_1(C_6) \sep C_8 = f_1(C_7) - C_6\comma \ed
  and find two relations:
  \be  f_1(C_5)  = C_5 \sep  f_1(C_8) = 2 C_8 \comma \ee
giving eight matrices and five relations in all.


 If we now form all commutators with ${\cal H}_0$, we 
find eight new matrices: 
\bd  C_9 =  f_0(C_5)  \sep C_{10} =  f_0(C_6)  
\sep C_{11} =   f_0(C_7)   \comma \ed
\bd  C_{12} =   f_0(C_{11}) \sep  C_{13} =   f_0(C_8)  
\sep C_{14} =   f_0(C_{13}) \comma \ed
\bd  C_{15}  =   f_0(C_{14})  - C_{13}  \sep C_{16} =   
f_0(C_{15})  \comma \ed
with four relations:
\be \label{last}
  f_0(C_{10})  = C_9 \sep  f_0(C_{9}) = C_{10}  
\sep   f_0(C_{12})  = C_{11} 
\sep  f_0(C_{16}) =  4 C_{15}  \comma \ee
 a total of 16 matrices and 9 relations.

At each stage we have a total of $2^m$ matrices (linearly 
independent provided
$L$ is sufficiently  large), satisfying a total of  
$1+ 2^{m-1} $ relations, 
for  $m = 1,2,3,4.$  Our numerical studies support 
the conjecture that this 
pattern continues  for all integers $m$ and all 
$N, L, r$ such that $0 < r < N$.




 \section{Reduced representation}


\setcounter{equation}{0}
Both $\ha$ and $\hb$ commute with the matrix 
\be R = X_1 X_2 \cdots X_L    \ee
which satisfies $R^N=1$ and has eigenvalues 
$1, \omega, \ldots, \omega^{N-1}$.   If 
\be \label{defvQ}
v_P\sd \eq N^{-1/2} \sum_{a=0}^{N-1} \omega^{- P a}
 \,  u_a  \ee
for $P = 0, 1, \ldots,N-1$, then
\be R \, v_P\sd = \omega^P v_P\sd \period \ee
The full $N^L$-dimensional space is the union of $N$ sub-spaces
$\cV_0, \cV_1, \ldots, \cV_{N-1}$ such that
\be R \, v \eq \omega^P v  \; \; \;  {\rm if} \; \; v \in 
\cV_P\sd \ee
and if two vectors $v, w$ belong to different sub-spaces, then
\be v^{\dagger} w \eq 0 \period \ee 
Clearly $v_P \sd \in \cV_P \sd$, and, because $\cal H$ commutes 
with $R$, 
\be  D U v_P\sd \in  \cV_P \sep D \, {\cS}_r \, U v_Q\sd \in 
\cV_P\sd \comma \ee
 where \be \label{PQr}
Q = P + r  \; \;  \;  ({\rm mod}  \; N) . \ee

{From} (\ref{defZW}) and (\ref{DUH}), $Z_a$ is a function of
$\alpha+\beta$, and $W_a(r)$ of $\alpha, \beta$ separately.
We define
 \be \label{defZQ}
\tZ_P (\alpha+\beta)  = \sum_{a=0}^{N-1} \omega^{P a} Z_a \sep 
  \tW_{PQ}(\alpha,\beta)   = \sum_{a=0}^{N-1} \omega^{P a} 
\, W_a(r)  \ee
and it  then  follows that
  \be  \label{ctZ}
\tZ _P (\alpha+\beta)  \eq v_P^{\dagger} \, D U  v_P\sd
  \sep
   \tW _{PQ}(\alpha,\beta )   \eq v_P^{\dagger} D \,
 {\cS}_r U  v_Q\sd  \comma   \ee
where $P, Q$ are again related by (\ref{PQr}).


The author observed\cite{Baxter1988} that if one pre-multiplies
the vector $v_P$ by various transfer matrices $T$ (in general 
with different values of the horizontal rapidity), then one does 
not generate the full vector space ${\cal V}_P$, but a smaller space
$V_P$ in which $T$ has  $2^m$ distinct eigenvalues, where
\be \label{defm}
m \eq m(P) \eq 
\left[  \frac{(N-1)L - P}{ N} \right]  \ee
and $[x]$ means the integer part of $x$.
Each eigenvalue occurs only once.

Label the basis vectors of $V_P$ by
\be s \eq \{ s_1, s_2, \ldots , s_m \} \comma \ee
where each $s_i$ takes the values $0$ or $1$. (We can 
think of each $ 1 - 2 s_i $ as an ``Ising spin", with value 
$\pm 1$.) Thus there are
$2^m$ vectors $\tilde{v}_s = 
\tilde{v}(s_1, s_2, \ldots , s_m )$, each
of dimension $N^L$. We define
\be \label{defkappa}
\kappa_s \eq s_1 + s_2 + \cdots + s_m \ee
so $ \kappa_s$ is an integer, and
\bd 0 \leq  \kappa_s \leq m \period \ed




In \cite{Baxter1989} we showed that we could choose the
vectors $\tilde{v}_s$ so that
\be \label{vQ}
v_P \eq \tilde{v}(0,0,\ldots ,0 ) \comma \ee
\bd
{\cal H}  \, \tilde{v}_s  \eq \mu_P \,  \tilde{v}_s \, -
N  \sum_{j=1}^m (1- k' \, \cos \theta_j) s_j  \, 
\tilde{v}_s  \ed
\be \label{Hv} 
  + N k' \sum_{j=1}^m \sin \theta_j \, 
{\tilde{v}} (s_1, \ldots ,-s_j , \ldots s_m) \comma \ee
where
\be \mu_P \eq [2 k' P + 
(1+k')(mN-NL+L)]  \, \period \ee

What we did not show, but believe to be true, is that
the vectors $\tilde{v}_s$ can all be chosen to be
independent of $k'$. For small $N, L$ we can generate
these vectors algebraically on the computer, and find this to be so.
This is consistent with the fact that $\cal H$ is 
linear in $k'$.\cite{NishinoDeguchi2006}


Define $2^m$ by $2^m$ matrices $\widehat{S}_j, \widehat{C}_j$ by
\be
(\widehat{S}_j)_{s,s'} \eq s_j \prod_{n=1}^m \delta(s_n,s'_n) 
\comma \ee
\be
(\widehat{C}_j)_{s,s'} \eq  \delta(s_j, 1 \! - \! s'_j) 
{\prod_{n=1}^m}^{\! \raisebox{-10pt}{*}}  \delta(s_n,s'_n) \comma \ee
where again the $*$ means that the term $n=j$ is excluded from the
product. Then from (\ref{Hv}), with respect to the basis 
vectors $\tilde{v}_s$, the hamiltonian
${\cal H}$ is now
\be \label{resH}
H \eq H_0 + k' H_1 \comma \ee
where
\be \label{defH0}
H_0 \eq L - N L + 2 N J_0\comma \ee
\be H_1 \eq 2 P + L - N L + 2 N J_1 \comma \ee
and \be \label{defJ0}
 2  J_0 \eq m I -  \sum_{j=1}^m \widehat{S}_j \comma \ee
\be \label{defJ1}
2 J_1 \eq m  I  + \sum_{j=1}^m (\cos \theta_j 
\, \widehat{S}_j + \sin \theta_j \, \widehat{C}_j )  
\comma  \ee
$I$ being the identity matrix of dimension $2^m$.
The reduced Hamiltonians $H, H_0, H_1, J_0,J_1 $ are also of 
dimension $2^m$.
If we replace $\ha, \hb$ in (\ref{defA}) by $H_0,H_1$, then again
we obtain the Onsager algebra (\ref{Onsalg}).

In this basis we see from (\ref{vQ}) that $v_P$ is replaced by
the $2^m$-dimensional vector ${\rm v}_P$ 
with entries
\ba ({\rm v} _P)_s = 1  \; \; \;  {\rm if } \; \; 
s &  =  & \{   0,0,\ldots ,0 \} 
\; , \; \; \\
 & = &  0     \; \; \; {\rm else.} \nonumber\ea
i.e.
\be \label{lastv}
{\rm v} _P = 
{ \scriptstyle  \left( \begin{array} {c} 1 \\ 0\end{array} 
\right) }
\otimes 
\left( \begin{array} {c} 1 \\ 0\end{array} \right) \, \otimes 
\;  \cdots  \; \otimes \, 
\left( \begin{array} {c} 1 \\ 0\end{array} \right) 
\period \ee

The vectors $\tilde{v}_s$ depend on $P$, so where necessary we 
write them as $\tilde{v}_s^P$. Similarly we may write 
$m, \theta_j, H,H_0,H_1$  as $m(P), \theta_j^P, H_P, 
H_{0}^{P}, H_{1}^{P}$. In particular, we consider two particular
values $P,Q$ of the index $P$, related by (\ref{PQr}),
 and set
\be m = m(P)  \sep \theta_i = \theta_i^P
 \;  \; ; \; \;  n = m(Q)  \sep {\theta'_j}= 
\theta_j^Q \comma  \ee
where $ i = 1, \ldots ,m$ and $j= 1, \ldots ,n$.

We have not yet defined the 
$\theta_1, \ldots , \theta_m$ (and $\theta'_1, \ldots , \theta'_{n}$). 
This is because we believe the equations of this paper 
to apply for {\em arbitrary } $\theta_1, \ldots , \theta_m$ and 
$\theta'_1, \ldots , \theta'_{n}$. We do not use the definitions 
here, but  for completeness they are are given in Appendix B.





\subsection*{Calculation of  $\tilde{Z}_P$, $\widetilde{W}_{PQ}$ }


The function $\tilde{Z}_P$  is unchanged if we replace 
${\cal H }, v_P$ in  (\ref{ctZ}), (\ref{DUH})  
by the reduced matrices and vectors $H, {\rm v}_P$.
The exponential $\e^{-\alpha H}$ is a direct product of
two-by-two matrices, so is easily calculated.
As in eqn. (3.16) of \cite{Baxter2008}, define functions
$\lambda (\theta), u(\alpha, \theta ),  v(\alpha, \theta ),
 w(\alpha, \theta )$
by
\be \label{deflambda}
\lambda (\theta) \eq \lambda \eq (1 - 2 k' 
\cos \theta + k'^2 )^{1/2} \comma  \ee
\bd u (\alpha, \theta ) \eq \cosh (N \alpha \lambda) \, + 
\, \frac{1-k' \cos \theta }{\lambda }
\sinh ( N \alpha \lambda) \ed
\be \label{defvqA}
v (\alpha, \theta ) \eq -\frac{k' \sin \theta}{\lambda} \; 
\sinh ( N \alpha  \lambda  )  \ee
\bd w (\alpha, \theta ) \eq \cosh (N \alpha \lambda) \, - 
\, \frac{1-k' \cos \theta }{\lambda }
\sinh ( N \alpha \lambda) \comma \ed
and let $U_j$ be the two-by-two matrix
\be U_j \eq 
\left( \begin{array}
{cc} u_{\q} (\alpha, \theta_j)  & v_{\q} (\alpha, \theta_j) \\
v_{\q} (\alpha, \theta_j)  & w_{\q} (\alpha, \theta_j) 
\end{array} \right)   \comma \ee
then
\be 
\e^{-\alpha { H}} \eq e^{- \mu_P \alpha } \,
U_1 \otimes U_2 \otimes \cdots \otimes U_m \period \ee
{From} (\ref{ctZ}), it follows that
\be \tilde{Z}_P (\alpha) \eq \e^{-\mu_P \alpha } 
\, u_P(\alpha, \theta_1 ) \cdots u_P(\alpha, \theta_m ) 
\period \ee

We can similarly write down an expression for 
of $\tilde{W}_{PQ}$, provided we replace $v_P$ by ${\rm v}_P$,
$v_Q$ by ${\rm v}_Q$,
the $\cal H$ in $D$ by 
$H_P$, the $\cal H$ in $U$ by  $H_Q$, and ${\cal S}_r$ by
a reduced matrix ${ S}_{PQ}$ with elements
\be \label{redS}
\left( {S}_{PQ} \right)_{s,s'} \eq (\tilde{v}_{s}^P)^{\dagger}
{\cal S}_r \tilde{v}_{s'}^Q \period \ee
Note that the set $s$ has $m$ entries, while $s'$ has $n$.
Hence the reduced matrix ${ S}_{PQ}$ is of dimension $2^{m}$ by 
$2^{n}$. It is not necessarily square.

Define
\be x_i \eq  \frac{v (\alpha, \theta_i ) }{ u (\alpha, \theta_i ) }
\sep x'_i \eq  \frac{v (\beta, \theta'_i ) }{ u (\beta, \theta'_i ) }
\period \ee
Then we obtain
\be \widetilde{W}_{PQ} (\alpha, \beta) \eq Z_P(\alpha) Z_Q	(\beta) 
\, {\cal D}_{PQ}\comma \ee
where
\be \label{defDPQ}
{\cal D}_{PQ} \eq \sum_{s{\vphantom'}} \sum_{s'} 
\, x_1^{s_1} x_2 ^{s_2} \cdots x_m^{s_m} 
\left( {S}_{PQ} \right)_{s,s'} {x'_1}^{s'_1} {x'_2} ^{s'_2} 
\cdots {x'_n}^{s'_n} \period \ee
However, this is still a $2^{m+n}$-dimensional summation.
In the following sections we firstly give an explicit 
conjecture for $\left( {S}_{PQ} \right)_{s,s'} $, and secondly
a conjectured expression  for ${\cal D}_{PQ}$ as
an $m$ by $m$ (or $n$ by $n$) determinant.
The formula (\ref{defDPQ}) is the same as eqn. (5.37) of 
\cite{Baxter2008}, but now  the
$\theta_j, \theta'_j$ are arbitrary.


\subsection*{The commutators}


Multiply any of 
the  equations (\ref{deff01}) - 
(\ref{last}) on the left by
the hermitian conjugate of an arbitrary  vector ${\tilde{v}_s^P}$ of 
the $P$-set, and on the 
right by a vector  ${\tilde{v}_{s'}^Q}$ of the $Q$-set. If we define
reduced matrices $C_1, \ldots , C_{16}$ analogously to (\ref{redS}),
then we see that (\ref{deff01}) - 
(\ref{last}) remain valid if we replace each $C_j$ by its reduced 
form, and any ${\cal H}_0$ to the 
left (right) of the $C$ matrix by $H_0^P$  ($H_0^Q$) and  
${\cal H}_1$ by $H_1^P$ ($H_1^Q$). 

We can use (\ref{defJ0}), (\ref{defJ1}) to replace 
$H_0^P, \ldots ,H_1^Q$ 
in these commutation relations by $J_0^P, \ldots ,J_1^Q$.
We have to take care to note that $0 \leq P,Q <N$
and $r = Q - P, \mod N$, so $0 < r <N$. The general
commutators (\ref{deff01}) become
\be  \label{redf01}
f_0(C)  \eq J_0^P C - C J_0^Q \sep 
 f_1(C)  \eq J_1^P C - C J_1^Q + \frac{1-\gamma}{2} C \comma \ee
where \be \label{defgamma}
\gamma = 1 \; \; {\rm if }\; \; P < Q \sep
\gamma = -1 \; \; {\rm if }\; \; P > Q  \period \ee

{From} (\ref{defm}), there are four possible cases to consider. We 
define a function $e(P,Q,i)$ in each case as follows.
\ba \label{restr}
1) \; \; e(P,Q,i)  \!\! \! \! \! & = \sin \theta_i  \;\;\; \;  
\; \; \; {\rm if} \; \; P < Q ,&  n  = m- 1, \;  \gamma = 1 \; , 
\nonumber \\
2)   \; \; \; \; \;\;\;\;\; \;\;\;\;& = \tan (\theta_i/2)  \;\;  
{\rm if} \; \;  P < Q , & n = m, \;\;\;\;\;\; \; \gamma = 1 \; , 
\nonumber \\
3)\; \;  \; \; \;\;\;\;\; \;\;\;\;&  = 1/\sin \theta_i  \;\;\; \;  
{\rm if} \;  \;  \; P > Q ,& n  = m+1,\; \gamma = -1\; ,\\
4)\; \; \;  \; \;\;\;\;\; \;\;\;\;  &  = \cot ( \theta_i/2)   
\;\; \;  {\rm if} \; \; P > Q ,&  n = m, \;\;\;\; \;\;\;\gamma = 
-1 \; . 
\nonumber  \ea
Similarly,   
\be e(Q,P,i) = 1/\sin \theta'_i ,  \; \cot(\theta'_i/2),
\;  \sin \theta'_i ,  \; \tan (\theta'_i/2)\ee
 for cases 1, \ldots , 4, respectively.

In the rest of this paper we take the $\theta_i, \theta'_i, x_i, 
y_i$ to be arbitrary and will no longer use
the relation  (\ref{defm}) between $N, L, P, m $, or between
$N, L, Q, n$. However, we stress that the restrictions
(\ref{restr}) appear to be necessary: in particular, 
we have not found any
generalizations to $n > m+1$ or $n < m-1$.




\subsection*{The reduced matrix ${S}_{PQ}$ }


Using (\ref{redf01}), (\ref{defgamma}), we
obtain two equations for  $S_{PQ}$, namely
\be \label{Comm0}
 J_0^P S_{PQ} = S_{PQ} \, J_0^Q \comma \ee
and
\be \label{Comm1} J_1^P J_1^P S_{PQ} 
-2 J_1^P S_{PQ} \, J_1^Q + S_{PQ} \, J_1^Q J_1^Q \eq 
  \gamma \,  (J_1^P S_{PQ} - S_{PQ} J_1^Q) 
\period \ee

These equations do not determine the normalization of $S_{PQ}$. To 
do this we note from (\ref{defS}),(\ref{defvQ}),(\ref{vQ}) that
\be \label{propS}
( S_{PQ} )_{s,s'} = 1 \; \; {\rm if} \;\; s = {\bf 0}\; \; 
{\rm and}  \; s'  = {\bf 0}' \period \ee
Here ${\bf 0} = \{ 0,0, \ldots , 0 \}$ has $m$ entries  and
 ${\bf 0 }' = \{ 0,0, \ldots , 0 \}$ has $n$ entries.

These give two commutation relations for $S_{PQ}$. The first is simple. 
{From} (\ref{defJ0}), $J_0^P$ is  a diagonal matrix with entries
\bd 0,1,2, \ldots ,m  \ed
and degeneracies $1, m, m(m-1)/2, \ldots $.
If we order the rows and columns of $J_0^P$ and $J_0^Q$ so that the 
diagonal entries are in increasing order, then (\ref{Comm0})
implies that $S_{PQ}$ is block-diagonal. 
More generally, (\ref{Comm0}) implies that
\be (S_{PQ})_{s,s'}  \eq 0  \; \; {\rm unless} \; \; \kappa_s =
 \kappa_{s'} \period  \ee

The second (double) commutation relation is more complicated, but 
algebraic computer calculations for small $m,n$ satisfying 
(\ref{restr})  strongly suggest that 
\vspace{1mm}

a) the relations (\ref{Comm0}) - (\ref{propS}) uniquely determine
$S_{PQ}$.

b)  the non-zero elements of $S_{PQ}$ are simple products.

\vspace{3mm}

\noindent To formulate our observations more specifically, we 
first need some 
further definitions. For a given set $s$, let $V$ be 
the set of integers $i$ such that
$s_i= 0$ and $W$ the set such that $s_i = 1$.
Hence, from (\ref{defkappa}), $V$ has $m-\kappa_s$ elements, while 
$W$  has $\kappa_s$.
Define $V'$, $W'$ similarly for the set $s'$. Set
\be c_i \eq \cos \theta_i \sep  c'_j \eq \cos \theta'_j 
\comma \ee
for $ 1 \leq i \leq m$ and  $ 1 \leq j \leq n$. Let
\ba \label{defADT}
A_{s,s'} = \prod_{  i \, \in\, W } \prod_{ j \, 
   \in V' } (c_i-c'_j)  &  , &
B_{s,s'} = \prod_{  i \, \in\, V } \prod_{ j \, 
   \in W' } (c_i-c'_j)\comma \nonumber \\
 C_{s}   \eq \prod_{  i \, \in\, W } \prod_{ j \, 
   \in V } (c_j-c_i)   & , &
D_{s'} \eq  \prod_{ \, i \, 
   \in V' } 
\prod_{  j \, \in\, W' \, }(c'_j-c'_i) \comma \\
 {\cal T} _{s} \eq \prod_{  i \, \in\, W } e(P,Q,i)
\; \;  & ,  &\;  
\; {\cal T}' _{s'} \eq \prod_{  i \, \in\, W' } 
e (Q,P, i)  \period 
\nonumber\ea
Then our calculations are consistent with the conjecture
\be \label{conjS}
(S_{PQ})_{s,s'} \eq 
\frac{  {\cal T} _{s} \,  {\cal T}' _{s'} \, A_{s,s'} B_{s,s'} }
{   C_s D_{s'}  }  \comma \ee
when $\kappa_s = \kappa_{s'}$, for all four cases  
(\ref{restr}).
This agrees with the symmetry
\be S_{PQ} = \left( S_{QP} \right)^{\dagger} \comma \ee
which follows from (\ref{defS}) and (\ref{redS}).  If we define
\be \label{defyy}
y_i \eq e(P,Q,i)  \, x_i \sep y'_i \eq e(Q,P,i) \, x'_i \comma \ee
it implies that (\ref{defDPQ}) can be written
\be \label{defDPQa}
{\cal D}_{PQ} \eq \sum_{s{\vphantom'}} \sum_{s'} 
\, y_1^{s_1} y_2 ^{s_2} \cdots y_m^{s_m} 
\left( \frac{  A_{s,s'} B_{s,s'} }
{   C_s D_{s'}  } \right) 
 {y'_1}^{s'_1} {y'_2} ^{s'_2} \cdots 
{y'_n}^{s'_n} \comma \ee
the sum being restricted to $s,s'$ such that 
$\kappa_s = \kappa_{s'}$.





\section{Determinantal conjecture}


\setcounter{equation}{0}

We emphasize that (\ref{conjS}) is independent of the definitions
(\ref{deftheta}) of the $\theta_i$ and $\theta'_i$, so should apply 
for arbitrary 
$\theta_i, \theta'_i$. In \cite{Baxter2008} we conjectured that
$\widetilde{W}_{PQ}$ could be written as a determinant, and this 
result also appears to be true for arbitrary 
$\theta_i, \theta'_i$. We repeat it here for this generalization.

Define two functions ${\cal{P}}_P(c), {\cal{P}}_Q(c)$ by
\be {\cal{P}}_{P} (c) =  \prod_{i=1}^m (c-\cos \theta_i) 
\sep {\cal{P}}_{Q} (c) =  \prod_{i=1}^{n} (c-\cos \theta'_i) 
\period \ee
They are polynomials in $c$, of degree $m, n$, respectively.
Let
\be 
\Delta_P (c) \eq \frac{\rm d}{{\rm d} c} \, {\cal{P}}_{P} (c) \ee
and similarly for $\Delta_Q (c)$. Let
\be
\epsilon \, (P, Q) = 1 \; \; {\rm if} \; {P}  <  Q \sep 
\epsilon \, (P,Q) = \; - 1 \; \; {\rm if} \; P  > Q  \ee
and define functions
\be \label{Bsumm}
f(P, Q, c) \eq \left[ \epsilon(P, Q)  \, {\cal{P}}_{Q}(c)/
\Delta_{P}  (c) \right]^{1/2} \comma \ee
\be {\cal B} (P,Q,c,c') \eq \frac{f(P,Q,c) \,  f(Q,P,c')}
{c-c'} \ee
for $P \neq Q$. They are rational functions of $c, c'$.

Let $B_{PQ}$ be the $m$ by $n$ matrix
with elements
\be
\left( B_{PQ} \right)_{ij} \eq {\cal B} (P,Q, \, \cos \theta_i ,
\, \cos \theta'_j) \period \ee
By construction it is orthogonal, in the sense that
\be B_{PQ}^T \, B_{PQ}^{\vphantom T} = I \; \; {\rm if}
\; \; m \geq n \sep
 B_{PQ}^{\vphantom T}\, B_{PQ}^T  = I \; \; {\rm if}
\; \; m \leq n   \period \ee

Define the $n$ by $m$ matrix $B_{QP}$ similarly, with $P, m,\theta_i$ 
interchanged with $Q, n,\theta'_i$, repectively.
Also define an  $m$ by $m$  diagonal matrix $Y_{PQ}$, and 
an $n$ by $n$ diagonal matrix  $Y_{QP}$, with elements
 \be 
(Y_{PQ} )_{i,j} \eq 
y_i \, \delta_{i,j}  \sep (Y_{QP} )_{i,j} \eq 
y'_i \, \delta_{i,j}  \period  \ee
We conjecture that
\be \label{conja}
{\cal D}_{PQ} \eq  \det [ I _m - 
Y_{PQ} \, B _{PQ} Y_{QP}
\, B_{QP }  ] \ee
or equivalently
\be \label{conjb}
{\cal D}_{PQ} \eq  \det [ I _{n} - 
Y_{QP}  \, B _{QP } Y_{PQ}
 \, B_{PQ}  ]  \period \ee
Here $I_m$ ($I_n$) is the identity matrix, of dimension $m$ ($n$). 

These equations (\ref{conja}),  (\ref{conjb}) are the same as
eqns.  (7.2), (7.3) of \cite{Baxter2008} when $\theta_i$, $\theta'_i$
are given as in Appendix B.


\section{Summary}
\setcounter{equation}{0}
If we consider the superintegrable chiral Potts model with
cylindrical boundary conditions, and fixed equal spins in
the top and bottom rows, we are led to the reduced hamiltonians
$J_0^P, J_1^P$ given by (\ref{defJ0}), (\ref{defJ1}).
The $\theta_i$ in  (\ref{defJ1}) are given as in Appendix B, but
for {\em all} $\theta_i$ it is true that if we take
\be A_0 = -4 J_1^P \sep  A_1 = 4 J_0^P \comma \ee
then we can define matrices $A_m,G_m$ such that the Onsager
algebra (\ref{Onsalg}) is satisfied.

To calculate the spontaneous magnetization we must introduce
the diagonal matrix ${\cal S}_r$ of (\ref{defS}). Its reduced form
$S_{PQ}$ of (\ref{redS}) satisfies the equations (\ref{Comm0}), 
(\ref{Comm1}).
Here we consider these equations for arbitrary $\theta_i$, 
$\theta'_i$ and conjecture that, together with the normalization
condition  (\ref{propS}), they uniquely define (\ref{redS}), and that
the solution is (\ref{conjS}). 

We show in \cite{Baxter2008}) that the spontaneous magnetization 
is given by an expression of the general form (\ref{defDPQ}).
Here we take the $x_i, x'_i$ therein to be arbitrary and define
related quantities $y_i, y'_i$ by (\ref{defyy}). We then generalize
our previous conjecture (7.2), (7.3) of \cite{Baxter2008}) to
(\ref{conja}), (\ref{conjb}), still keeping the $\theta_i, \theta'_i$ 
arbitrary (but note that $m, n$ must satisfy the restictions 
(\ref{restr}).

The factors $  {\cal T} _{s},   {\cal T}' _{s'} $ can be removed from 
the equations 
 (\ref{Comm0}),  (\ref{Comm1}) by incorporating them into the $J_0$,
$J_1$  expressions in (\ref{conja}), (\ref{conjb}). We do this in 
Appendix C.   Our conjectures 
then reduce to rational identities in the arbitrary variables 
$c_i, c'_i$. In this form they should be easier to 
establish.




\section{Acknowledgements}
The author thanks Helen Au-Yang and Jacques Perk for very helpful
and encouraging discuusions, and the Rockefeller Foundation for a 
residency at the Bellagio Centre, where this paper
was written. He has great pleasure in submitting this paper for this
special issue of the Journal of Statistical Physics  and thanks Joel 
Lebowitz and his colleagues for the inspiration provided by the
50 years of Yeshiva/Rutgers meetings.




\section*{Appendix A}


\renewcommand{\theequation}{A\arabic{equation}}
\appendix
\setcounter{equation}{0}

Here we prove the commutation relation  (\ref{eqforC2}).
We take $0 < r <  N$.

Since ${\cS}_r = Z_1^{\, r}$ commutes with all the terms in the 
definition (\ref{defhh}) of 
$\hb$ except the $j=1$ term, we can replace $\hb$ in (\ref{defhh}) by
\be {\cal H}_1   \eq  - 2 \sum_{n=1}^{N-1} 
\frac{X_1^n}{1-\omega^{-n} } \period \ee


Also, the matrices 
$Z_1, X_1$ defined by (\ref{ZX}) satisfy
\be  Z_1 X_1 \eq \omega X_1 Z_1 \period \ee
This relation is unchanged if we replace $Z_1, X_1$ 
by $X_1^{-1}, Z_1$, and indeed there is a similarity transformation
that does this. Doing this and using the formula (\ref{formula}), 
it follows  that for the purposes of this Appendix we can take
$\hb$ to be the $N$ by $N$ diagonal matrix
\be
\hb  \eq \left( \begin{array} {c c c  c c } 1-N & 0 &  & \cdots & 
0 \\ 0 & 3-N & &  \cdots  & 0 \\
0 & 0 & & \cdots   & 0 \\
\cdots  & \cdots   & & \cdots  & \cdots \\
0 & \cdots &  & 0 & N-1  \end{array} \right) \ee
and  ${\cS}_r$  to be  the matrix whose elements $(i,j)$ are  
zero unless
$j = i+r$ (mod $N$) when they are one. We can therefore
write ${\cS}_r$ as
\be {\cS}_r = A + B \comma \ee
where $A, B$ are the $N$ by $N$ matrices

\be  A \eq \left( \begin{array} {c  c } {\bf 0} & {\bf 1}   \\
{\bf 0}  & {\bf 0}    \end{array} \right) \sep   B  \eq 
\left( \begin{array} {c  c } {\bf 0} & {\bf 0}   \\
{\bf 1}  & {\bf 0}    \end{array} \right) \ee
the $ {\bf 1} $ in the equation for $A$ being the identity matrix 
of dimension $N-r$
and the  $ {\bf 1} $ for $B$ being of dimension $r$. All other 
elements of  $A$ and $B$ are zero.

We readily see that
\be 
[\hb,A] = - 2 r A \sep [\hb,B] = 2 (N-r) B  \period \ee

Hence
\be [\hb,{\cS}_r]+2 r {\cS}_r = 2 N B \sep [\hb,B]+2 r B = 2 N B  
\period \ee
These relations are of course independent of similarity 
transformations. 
Setting $C_1 = {\cS}_r$ and $ C_2 = B$, we see that we have 
proved the relation  (\ref{eqforC2}).




\section*{Appendix B}


\renewcommand{\theequation}{B\arabic{equation}}
\appendix
\setcounter{equation}{0}

For a given value of $P$ with $0 \leq P < N$, define a polynomial 
$\rho (w)$, of  degree $m$, by
\be \label{defP}
 \rho(z^N) \eq z^{-{P} } \, \sum_{n=0}^{N-1} 
\omega^{(L+{P} )n} {(z^N-1)^L/(z-\omega^n)}^L \period \ee
Let its zeros be $w_1, \ldots ,w_m$ and define 
$\theta_1, \ldots ,\theta_m$  by
\be \label{deftheta}
\cos \theta_j \eq  (1+w_j)/(1-w_j) \sep 0 
< \theta_i < \pi  \comma \ee
for $j = 1,\ldots, m $. These are the $\theta$'s of the 
superintegrable chiral Potts model.\cite{Baxter2008, Baxter1988} 
They depend on $L, N, P$, so we may
write $\theta_i$ as $\theta_{P,i}$. They are independent
of $k'$. We do {\em not} use them in this paper. In particular our 
conjectures  (\ref{conjS}), (\ref{conja}),
 (\ref{conjb}) are  for  arbitrary $\theta$'s.


\section*{Appendix C}


\renewcommand{\theequation}{C\arabic{equation}}
\appendix
\setcounter{equation}{0}

Here we explicitly write the commutation relations
(\ref{Comm0}), (\ref{Comm1}) in terms of matrices that
are rational functions of $c_i = \cos \theta_i$, 
$c'_i = \cos \theta'_i$.

{From}  (\ref{defADT}), we are led to define a modified
matrix $\widetilde{S}_{PQ}$ by  the equivalence transformation
\be \label{trans}
S_{PQ} \eq E_{PQ} \widetilde{S}_{PQ} E_{QP} \comma \ee
where  $E_{PQ}$ is a direct product of $m$ two-by two diagonal 
matrices:
\be E_{PQ} \eq  
{ \scriptstyle  \left( \begin{array} {cc} 1 & 0 \\ 0  & e_1 

\end{array} 
\right) }
\otimes 
\left( \begin{array} {cc} 1 & 0 \\ 0 & e_2 \end{array} \right) \, 
\otimes  \;  \cdots  \; \otimes \, 
\left( \begin{array} {cc} 1 & 0 \\ 0 &  e_m\end{array} \right) 
\comma \ee
and $e_i = e(P,Q,i)$. The matrix $E_{QP}$ is defined similarly,
with $m$ replaced by $n$ and $e_i$ replaced by $e'_i = e(Q,P,i)$.
We also define   $ {\widetilde{J}_1}^{\, P},
{\widetilde{J}_1}^{\, Q}$ by\footnote{
Each actually depends on both $P$ and $Q$ because of the restrictions
(\ref{restr}).}
\be J_1^P = E_{PQ}  \, {\widetilde{J}_1}^{ \, P} \, E_{PQ}^{-1} \sep 
J_1^Q = E_{QP}^{-1}  \, {\widetilde{J}_1}^{\,  Q} \, E_{QP} 
\period \ee


 For the four cases (\ref{restr}), let
\be \label{defxi}
\xi_i \eq  1-c_i^2 , \; \; 1-c_i  \; \; 1, \;\; 1+c_i\comma \ee
repectively, and set
\be \Gamma_i \eq {\rm e} \otimes \cdots \otimes 
\left( \begin{array} {cc} 0 & \; \xi_i \\ (1-c_i^2)/\xi_i & \; 0
 \end{array} \right) \otimes  \cdots \otimes {\rm e}  \comma \ee
each ${\rm e}$ being the two-by-two identity matrix and the displayed
matrix  being in position $i$. Then
\be \label{defJt1}
2 {\widetilde{J}_1}^{\, P} \eq m  I  + \sum_{j=1}^m (c_j 
\, \widehat{S}_j +\Gamma_j )  
\period \ee
It is a polynomial in $c_1, \ldots , c_m$.
The matrix ${\widetilde{J}_1}^{\, Q}$ is also given by (\ref{defxi}) -
 (\ref{defJt1}), but with $m, c_i$ replaced by $n, c'_i$.

With these equivalence and similarity transformations, the 
 commutation relations (\ref{Comm0}), (\ref{Comm1}) become
\be \label{Comm0a}
 J_0^P  \widetilde{S}_{PQ}= \widetilde{S}_{PQ}\, J_0^Q \comma \ee
and
\bd \label{Comm1a} {\widetilde{J}_1}^{\, P} 
{\widetilde{J}_1}^{\, P} \widetilde{S}_{PQ} 
-2 {\widetilde{J}_1}^{\, P} \widetilde{S}_{PQ} \, 
{\widetilde{J}_1}^{\, Q} + \widetilde{S}_{PQ} \,
{\widetilde{J}_1}^{\, Q} {\widetilde{J}_1}^{\, Q} \eq 
 \gamma \,  ({\widetilde{J}_1}^{\, P} \widetilde{S}_{PQ} - 
\widetilde{S}_{PQ} {\widetilde{J}_1}^{\, Q}) 
 \ed
and, from (\ref{conjS}) and (\ref{trans}), our conjectured solution 
is
\be 
\left(  \widetilde{S}_{PQ} \right)_{s,s'} \eq
\frac{  A_{s,s'} B_{s,s'} }
{   C_s D_{s'}  }  \; \delta (\kappa_s , \kappa_{s'}) \period  \ee


 \end{document}